\documentclass{PoS}

\usepackage{psfig}

\def\la{\ifmmode\stackrel{<}{_{\sim}}\else$\stackrel{<}{_{\sim}}$\fi} 
\def\ga{\ifmmode\stackrel{>}{_{\sim}}\else$\stackrel{>}{_{\sim}}$\fi} 

\title{Revealing Cosmic Magnetism with Radio Polarimetry}

\ShortTitle{Revealing Cosmic Magnetism with Radio Polarimetry}

\author{{Bryan M.\ Gaensler}%
         \thanks{Australian Research Council Federation Fellow}\\
        School of Physics, The University of Sydney\\
        E-mail: \email{bgaensler@usyd.edu.au}}

\abstract{While gravitation sustains the on-going evolution of the
cosmos, it is magnetism that breaks gravity's symmetry and that provides
the pathway to the non-thermal Universe.  By enabling processes such as
anisotropic pressure support, particle acceleration, and jet collimation,
magnetism has for billions of years regulated the feedback vital
for returning matter to the interstellar and intergalactic medium.
After reviewing recent results that demonstrate the unique view of
magnetic fields provided by radio astronomy, I explain how the
Square Kilometre Array will provide
data that will reveal what cosmic magnets look like, how they formed,
and what role they have played in the evolving Universe.}

\FullConference{From Planets to Dark Energy: the Modern Radio Universe\\
		 October 1-5 2007\\
		 The University of Manchester, UK}

\begin{document}

\section{Introduction: Magnetism Matters!}

Most astronomical research does not explicitly incorporate the
effects of magnetic fields. Nevertheless, there are fundamental
reasons why we need to try to understand the various roles played
by magnetism in astrophysical processes. 

Zweibel \cite{zwe06}
provides two specific, clear motivations for further studies of
cosmic magnetic fields.
First and foremost, the origin of magnetic fields in the Universe is
a fundamental and unsolved cosmological problem \cite{gr01,wid02}.
We do not know whether the first magnetic fields emerged through
exotic processes such as phase transitions or string cosmology, or
through standard plasma physics such as turbulence, instabilities or the
battery effect. It is unclear whether the diffusion of these seed fields
throughout the Universe was then a top-down or a bottom-up process. Were
magnetic fields in the early Universe strong enough to moderate the
formation of large-scale structure?  What role could these fields have
played in the formation of the first stars and galaxies?

Second, magnetic fields are the key to answering many long-standing
problems in plasma physics and astrophysics \cite{mel80b,gll05}.  On the
largest scales, the coherent magnetic fields that stretch over enormous
physical scales  in galaxies and clusters allow us to test the extremes
of dynamo theory and turbulence. Magnetic fields are at the core of
any viable theory for the acceleration and propagation of cosmic rays.
Magnetism drives the physics, geometry and evolution of active galactic
nuclei (AGN). And finally, a whole variety of important physical processes
that drive galactic ecology, including star formation, thermal conduction,
diffusion and accretion, all rely heavily on the strength and geometry
of the ambient magnetic field.

If our goal as astronomers is to better understand the Universe,
then we need to include magnetic fields in our observations and in
our models. In this paper, I outline some of the recent progress
and future prospects in this area. In \S\ref{sec_map}, 
I briefly review the ways in which radio data can map magnetic
fields out to large distances. In \S\ref{sec_mw} \& \S\ref{sec_lmc},
I show some recent applications to the large-scale magnetic fields
of the Milky Way and of the Large Magellanic Cloud (LMC), respectively.
In \S\ref{sec_ska}, I discuss some of the experiments on magnetic
fields at intermediate and high redshifts that can be conducted
with the Square Kilometre Array (SKA).

\section{Mapping Magnetic Fields}
\label{sec_map}

There are various indirect approaches to studying magnetic fields in
astrophysical sources \cite{zh97}. Measurements of optical starlight
polarisation (e.g., \cite{hei96}), polarised radio synchrotron emission
(e.g., \cite{bh96}) and infrared/sub-mm dust polarisation (e.g.,
\cite{wkc+00}) all characterise the structure of the magnetic field in the
plane of the sky, $B_\perp$.  While this provides information on two of
the three spatial components of the field vector, a crucial limitation
of these approaches is that they all only provide the orientation of $B$, but
not its direction. For example, simple compression of a tangled magnetic
field can produce a set of polarisation vectors which will appear quite
ordered, but which have no spatial coherence.

Zeeman splitting is distinct from these other approaches, in
that it measures the line-of-sight component of the magnetic field
($B_\parallel$), and can determine the direction of this field (e.g.,
\cite{th86,rs90}).  However, most Zeeman experiments require long
integrations, and often probe localised regions of relatively high
gas density.

\begin{figure}[b!]
\centerline{\psfig{file=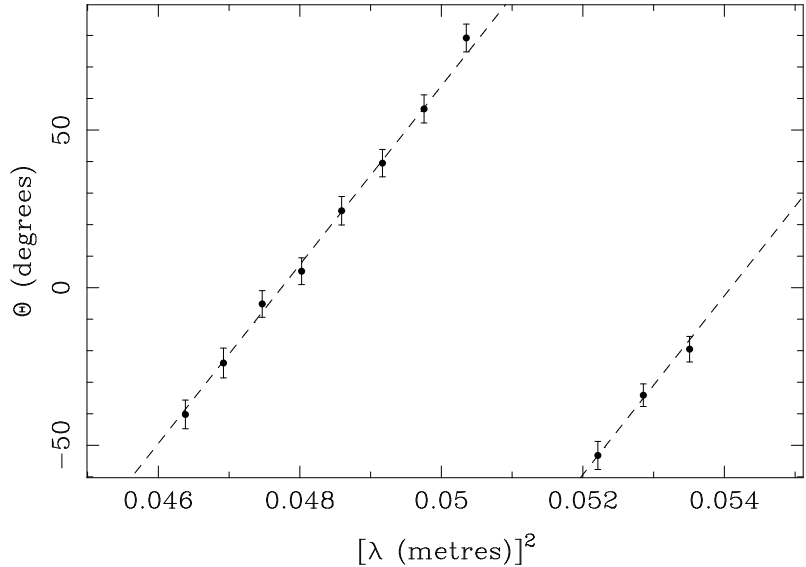,width=0.7\textwidth,clip=,angle=270}}
\caption{Faraday rotation in the Galactic interstellar medium, seen
toward the radio pulsar B1154--62 \protect\cite{gmg98}. The data
points indicate multi-channel spectropolarimetric measurements in
the 20~cm band made using the Australia Telescope Compact Array
(ATCA), while the dashed line shows the best fit of
Equation~(\protect\ref{eqn_rm1}) to these data. The slope of this line
is RM~$=+495\pm6$~rad~m$^{-2}$, indicating that the mean
magnetic field along this sightline is directed toward us.}
\label{fig_psr}
\end{figure}

The remaining approach to studying magnetic fields is {\em Faraday rotation},
which has proven to be a very powerful probe of $B_\parallel$. Faraday
rotation is a change in the linear polarisation angle of a radio signal as it
propagates through a magnetised plasma. The orientation of the 
observed electric field vector is:
\begin{equation}
\Theta = \Theta_0 + {\rm RM}~\lambda^2,
\label{eqn_rm1}
\end{equation}
where $\lambda$ is the observing wavelength, $\Theta_0$ is the intrinsic
polarisation angle emitted by the source, and $\Theta$ is the observed
polarisation angle. The rotation measure (RM) is a path integral through the
magneto-ionised foreground:
\begin{equation}
{\rm RM}~= K \int n_e B_\parallel dl
\label{eqn_rm2}
\end{equation}
where the free electron density, $n_e$, and line-of-sight field strength,
$B_\parallel$, are integrated from the observer to the source along a line
element $dl$, and $K$ is a constant. An example of Faraday rotation seen in
the polarised emission from a radio pulsar is shown in Figure~\ref{fig_psr}.

Faraday rotation has three important advantages. First, as for Zeeman
splitting, it provides a direction for $B_\parallel$ along a given
sightline, allowing us to test for field coherence if we have 
RM data for several adjacent positions on the sky. Second, since Faraday
effects are strongest in the radio part of the spectrum, interstellar
extinction can be disregarded, and magnetic fields can thus be probed out
to cosmological distances. Finally, Faraday rotation is an ``absorption''
experiment, in the sense that the signal-to-noise ratio of the measurement
is determined by the flux of the polarised background source, rather than
any properties of the region being studied. Measurements of magnetic
fields in regions that are not easily observable directly, such as the
Galactic halo and the intergalactic medium (IGM), then become possible,
provided that appropriate background sources can be identified.

\section{The Galactic Magnetic Field}
\label{sec_mw}

It has long been established that there are magnetic fields of significant
strength throughout the interstellar medium of the Milky Way
\cite{fer49,dg49}. Points of general consensus are that the field
has both large-scale (ordered) and small-scale (random) components,
that the magnetic field is concentrated in and is generally oriented
parallel to the disk of the Galaxy, and that the ordered component
of the field probably broadly follows the spiral arms. Studies of pulsar
RMs, extragalactic RMs and optical starlight polarisation show that
the large-scale field within a kpc of the sun is directed clockwise
(as viewed from the North Galactic Pole) \cite{hei96,man72}, but that
the field in the Sagittarius-Carina spiral arm, a few kpc closer to
the Galactic Centre, is oriented counterclockwise \cite{sk80,tn80}.
This clearly indicates that a large-scale {\em reversal}\ of the
large-scale magnetic field occurs somewhere between these two arms.
The presence and properties of such reversals provide crucial constraints
on dynamo theories and on the origin of galactic magnetic fields
(see \cite{shu05}).

\begin{figure}[t!]
\centerline{\psfig{file=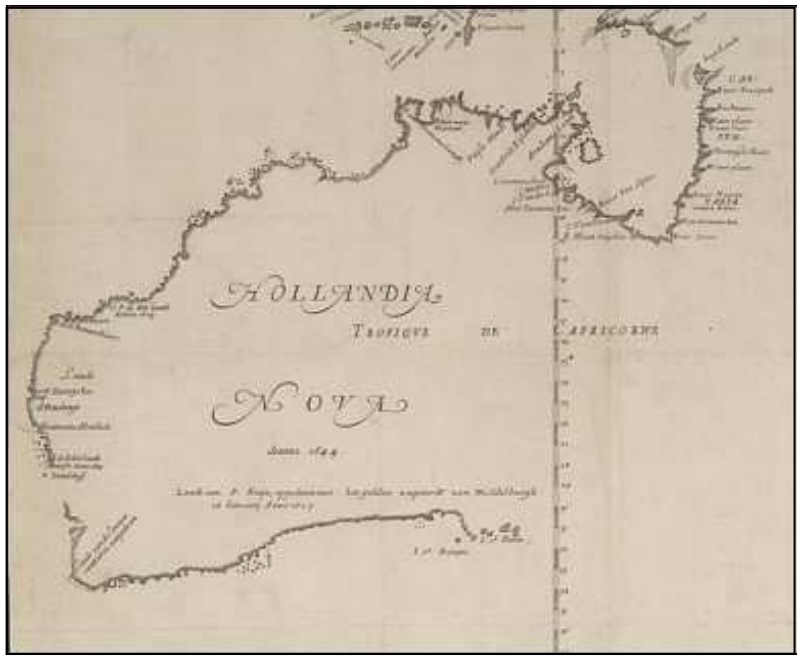,width=0.7\textwidth}}
\caption{An early map of Australia, drawn by the French cartographer
Melchis\'edech Th\'evenot
(``\protect\href{http://nla.gov.au/nla.map-rm689a}{Hollandia Nova
detecta 1644; Terre Australe decouuerte l'an 1644}'', Map RM 689A,
\cite{the63}).  Reproduced with permission of the National Library
of Australia.}
\label{fig_thevenot}
\end{figure}

For more distant parts of the Galaxy, especially beyond the Galactic
Centre, the geometry of the field is still unclear, mainly due to
the lack of known pulsars in these regions. Recent studies have
come to a variety of conclusions as to the overall Galactic field
geometry. Just to provide a few examples: Weisberg et al.\ \cite{wck+04}
have argued that the field follows the spiral arms, with a reversal
between each arm; Han et al.\ \cite{hml+06} have proposed that the
field is a spiral, but with reversals on both sides of every spiral
arm, so that fields in arm and inter-arm regions are directed in
opposite directions; Vall\'ee \cite{val05} has suggested that the
field is purely azimuthal rather than a spiral, with reversals in
concentric rings. As noted by Vall\'ee \cite{val02}, the situation
is not unlike the early exploration of Australia, during which half
the coastline had been mapped in detail, with the more distant parts
trailing off into ``terra incognita'', as shown in
Figure~\ref{fig_thevenot}.


\subsection{Polarised Extragalactic Sources}
\label{sec_xgal}

Most of the work mentioned above has relied heavily on pulsar RMs.
Pulsars have several advantages for such studies: they are at known
distances (albeit with significant uncertainty for individual sources),
they are within the Galaxy (so that pulsars at a range of distances
can be used to make differential magnetic field measurements), and both
dispersion and Faraday rotation of their signals can be measured (so that
the $n_e dl$ term in Eqn.~[\ref{eqn_rm2}]) can be independently
estimated).

However, pulsars also have their limitations. Most pulsars are
relatively nearby. This means that individual H\,{\sc ii}\ regions
and other dense gas clumps along the line of sight can make a large
contribution to the overall RM, making it more difficult to probe
the properties of large-scale magnetic fields (e.g.,  \cite{mwkj03}).
The distance estimates to individual pulsars from standard models
(e.g., \cite{cl02}) can sometimes be significantly in error, making
it difficult to match the magnetic field direction inferred from a
pulsar's RM to Galactic structure such as spiral arms. Pulsars are
a small population of faint sources (there are less than 2000 known
pulsars, with a median 1.4~GHz flux of $\sim0.5$~mJy), so that the
density and total number of RM sightlines that they can provide are
both relatively low.  Finally, even in regions such as along the
Galactic plane, where the sky density of pulsar RMs is higher than
average, the RM sample cannot be easily smoothed to bring out
large-scale structure, because each pulsar is at a different distance.

To make substantial further progress in studying the Milky Way's
magnetism, an additional approach is therefore needed.  We have
consequently undertaken a coordinated effort to greatly expand the
number of polarisation and RM measurements for extragalactic sources
(i.e., radio galaxies and AGN). These sources are at large distances,
and so provide a measurement of the Faraday rotation through the
entire Milky Way along a particular direction. In fact, the observed
RM is the sum of multiple contributions: from the source itself,
from the IGM, from the Earth's ionosphere, and from the Milky Way.
However, in most cases, and certainly when averaged over a large
sample of such data, the first three terms are small and can be
disregarded, and the total RM can be used as a useful probe of the
Galaxy's magnetism.

It is important to emphasise that RM data for extragalactic
sources only provides an integral of $n_e B_\parallel dl$, as per
Equation~(\ref{eqn_rm2}). Thus if the actual value of $B_\parallel$ is
of interest, a model is needed of $n_e$ as a function of $l$ (e.g.,
\cite{cl02}).  However,
many important problems require knowledge only of the geometry of $B$,
without accurate estimates of its strength.  Since the sign of the
RM is determined by the sign of $B_\parallel$, the overall magnetic
configuration can be studied from the RMs alone, without needing to
invert the integral.

\begin{figure}
\centerline{\psfig{file=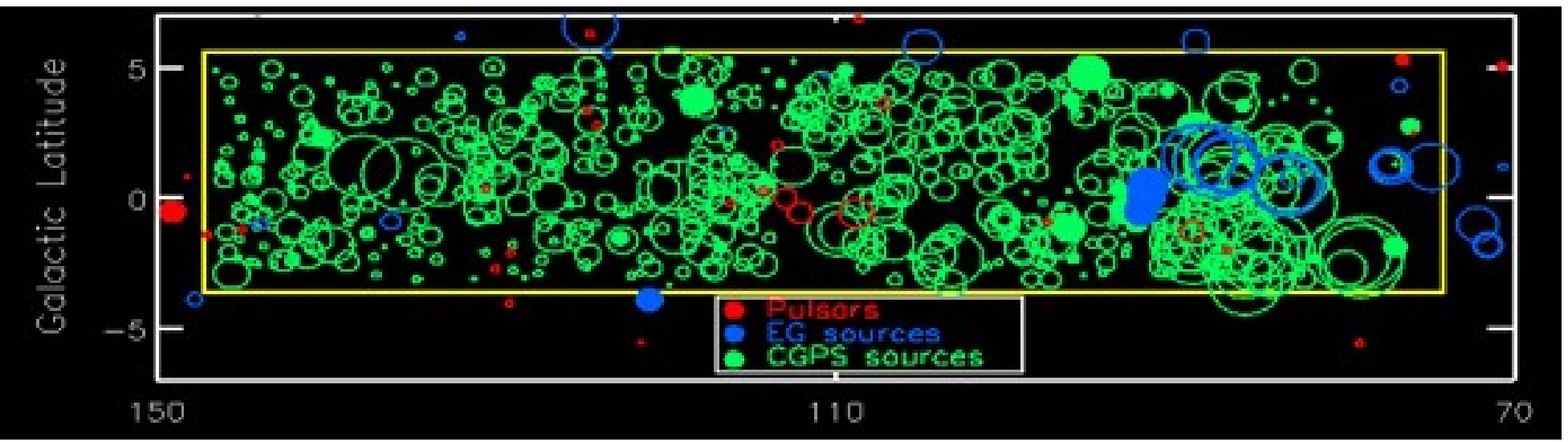,width=1.0\textwidth}}
\medskip
\centerline{\psfig{file=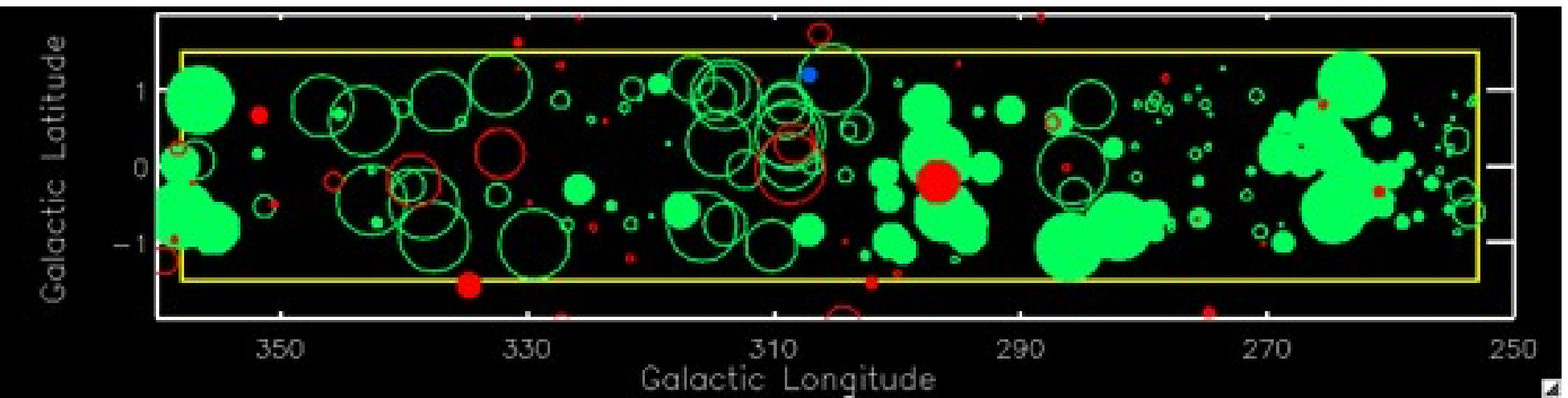,width=1.0\textwidth}}
\caption{Distribution of RMs in two strips along the
Galactic plane, corresponding to the survey regions
of the CGPS (upper panel) and SGPS (lower panel) \cite{btj03,bhg+07}.
The circles mark the positions of RM measurements: blue for previously
measured extragalactic RMs, red for pulsar RMs, and green for new
CGPS/SGPS data. The radius of each circle corresponds to the magnitude
of the RM for that source; filled circles indicate positive RMs and
open circles show negative RMs.  Figures are courtesy of Jo-Anne Brown.}
\label{fig_rms}
\end{figure}

Brown et al.\ \cite{btj03,bhg+07} have recently completed two large
surveys of extragalactic RMs behind the Galactic plane: 380 RMs in
the outer Galaxy, derived from the Canadian Galactic Plane Survey
(CGPS)\footnote{\href{http://www.ras.ucalgary.ca/CGPS/}{http://www.ras.ucalgary.ca/CGPS/}}
and 148 RMs in the inner Galaxy, measured as part of the Southern
Galactic Plane Survey
(SGPS).\footnote{\href{http://www.atnf.csiro.au/research/HI/sgps/}{http://www.atnf.csiro.au/research/HI/sgps/}}
The results are shown in Figure~\ref{fig_rms}; the vast improvement
in the sample size compared to previous data-sets reveals a striking
coherence in the large-scale magnetic field configuration. For the
CGPS, the RMs are almost all negative, showing that the Perseus
spiral arm (through which these sources are viewed) has a clockwise
magnetic field \cite{bt01}. For the SGPS, the situation is considerably
more complicated, with alternating regions of positive and negative
RMs, with regions of low $|{\rm RM}|$ between them.  The implications
of the RM structure seen in the SGPS are considered in the following
section.

\subsection{The Magnetic Geometry of the Milky Way}

\begin{figure}
\centerline{\psfig{file=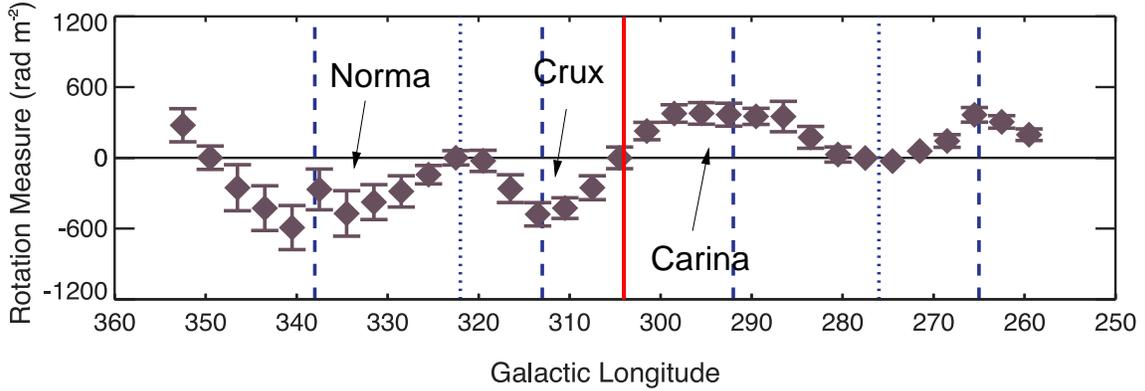,width=\textwidth,clip=}}
\caption{RM vs.\ Galactic longitude for extragalactic RMs in the
SGPS region. The purple diamonds represent smoothed RM data,
with error bars indicating the standard error of the mean in each bin.
The dotted (dashed) lines indicate approximate longitudes of minimum
(maximum) $|{\rm RM}|$ in the smoothed distribution. The red line marks the
approximate transition from predominantly positive RMs ($\ell \la 304^\circ$)
to predominantly negative RMs ($\ell \ga 304^\circ$).
Adapted from \cite{bhg+07}.}
\label{fig_sgps}
\end{figure}

Figure~\ref{fig_sgps} shows a smoothed version of the extragalactic
RM data from the lower panel of Figure~\ref{fig_rms}. The spatial
coherence of Faraday rotation as a function of Galactic longitude
is very clear.  There are local maxima of $|{\rm RM}|$ in three directions:
$\ell \approx 292^\circ$, $\ell \approx 312^\circ$ and $\ell \approx
338^\circ$, as shown by the dashed vertical lines in Figure~\ref{fig_sgps}.
These sightlines are tangent to the Carina, Crux and Norma spiral
arms, respectively. On either side of these peaks are directions
where $|{\rm RM}| \approx 0$~rad~m$^{-2}$, shown by dotted lines
in Figure~\ref{fig_sgps}. Superimposed on this oscillatory behaviour,
is an overall large-scale signature: for $\ell \la 304^\circ$, RMs
are almost all positive, and for $\ell \ga 304^\circ$, RMs are
predominantly negative (apart from some high RMs close to the
Galactic Centre). This dividing line is marked in red in
Figure~\ref{fig_sgps}.

Qualitatively, these patterns in RM immediately reveals some
properties of the global magnetic field structure of the Galaxy.
The peaks in $|{\rm RM}|$ along each spiral arm demonstrate that
the field strength and/or gas density are high in these regions,
while the low values of $|{\rm RM}|$ indicate low values of these quantities
in the inter-arms (see \cite{bhg+07} for a detailed discussion).
The change in overall sign of RM at $\ell \approx 304^\circ$ suggests
the presence of a large scale reversal in the field along this sightline.
of this direction.

\begin{figure}[b!]
\vspace{5cm}
\caption{A simple model of the magnetic field in the southern Galaxy,
from a joint fit to the RMs of 149 extragalactic sources and 120
pulsars. Coloured regions show the total strength of the model
magnetic field in separate regions bounded by the green lines.  The
limiting longitudes of the SGPS are shown as black lines, while the
grey scale represents the NE2001 electron density model of Cordes
\& Lazio \cite{cl02}.  Adapted from \cite{bhg+07}.}
\label{fig_model}
\end{figure}

\begin{figure}[b!]
\centerline{\psfig{file=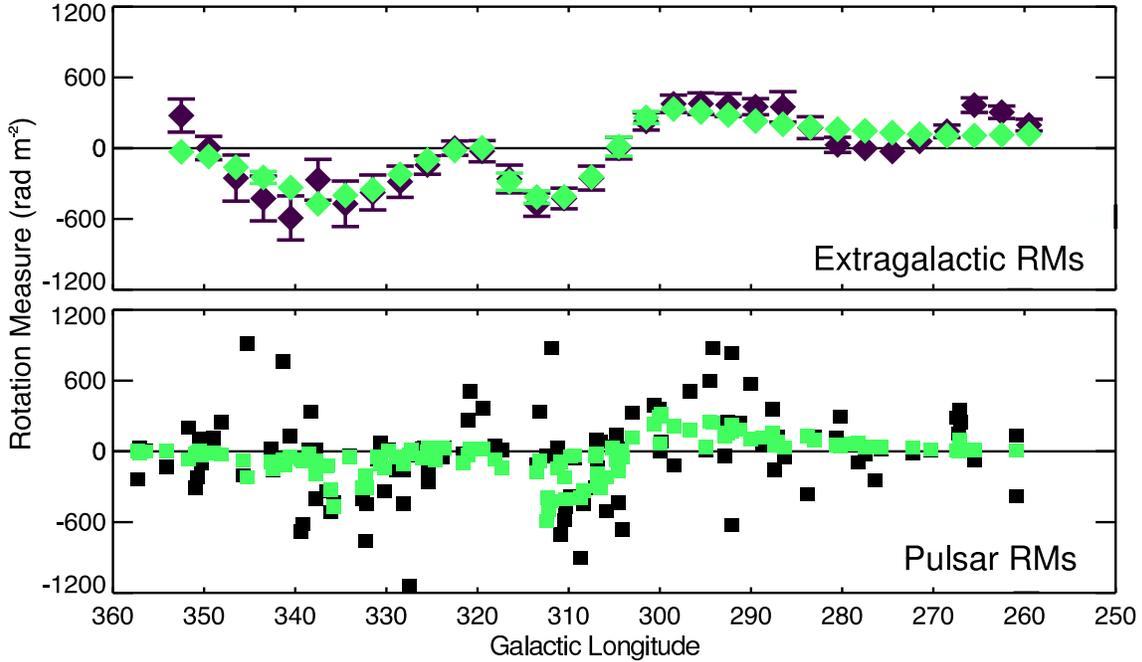,width=\textwidth}}
\caption{RM vs.\ Galactic longitude for extragalactic sources (upper;
purple points) and pulsars (lower; black points) in the SGPS region;
the extragalactic data have been smoothed as in
Fig.~\protect\ref{fig_sgps}.  The green symbols show the RMs predicted
by the best-fit model of \protect\cite{bhg+07} at the positions of
each of the observed sources. The model data for extragalactic
sources have been smoothed in the same way as for the observations.
Adapted from \protect\cite{bhg+07}.}
\label{fig_fit}
\end{figure}

This is quantified in Figure~\ref{fig_model}, where we show a joint
fit to extragalactic and pulsar RMs in the southern Galaxy, allowing
a series of concentric, spiral, annuli to each have a differing
field strength and field direction.
The best fit shows that the Galactic magnetic field is
primarily clockwise, except for a strong counterclockwise field in
the Scutum-Crux spiral arm (and possibly also in the molecular ring
in the inner Galaxy).  Sightlines at $\ell > 304^\circ$ pass through
the Scutum-Crux arm, and are dominated by negative RMs; at $\ell < 304^\circ$,
field lines are directed toward the observer, and RMs are consequently
positive.  The quality of the fit is indicated in Figure~\ref{fig_fit},
where we compare RM data to the predictions of the model in
Figure~\ref{fig_model}.  For extragalactic data, the model RMs and
the data match very well.  For pulsars, the scatter is larger (mainly
because pulsar RM data cannot be meaningfully smoothed), but the
major features are reproduced.

This global fit is at odds with earlier studies utilising smaller
data-sets, in that it suggests that the Galaxy can be modelled with
a predominantly clockwise field, plus a single reversed region.
This structure is in line with what is seen also for other spiral
galaxies, but needs to be verified by a better mapping of extragalactic
RMs in the first Galactic quadrant.

\section{The Magnetic Field of the Large Magellanic Cloud}
\label{sec_lmc}

\begin{figure}
\centerline{\psfig{file=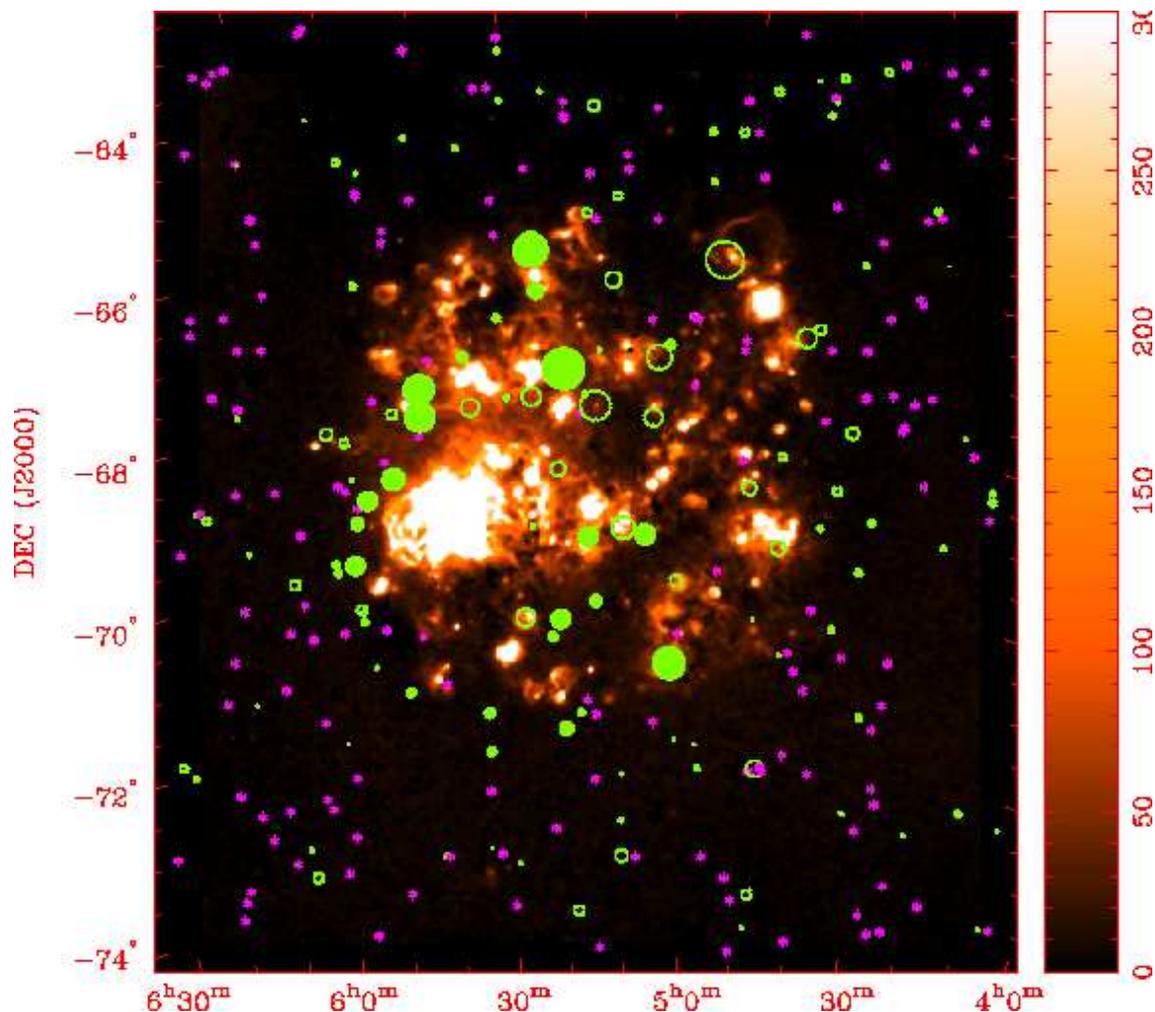,width=\textwidth}}
\caption{RMs of extragalactic sources behind the Large Magellanic
Cloud \cite{ghs+05}.  The image shows the distribution of emission measure toward
the LMC in units of pc~cm$^{-6}$. The symbols show the position,
sign and magnitude of extragalactic RMs: filled (open) circles
correspond to positive (negative) RMs, while asterisks indicate RMs
which are consistent with zero within their errors.  The diameter
of each circle is proportional to the magnitude of the RM.}
\label{fig_lmc}
\end{figure}

The LMC is also particularly amenable to extragalactic RMs as a
probe of its magnetic field, because of its large angular extent
($\sim6^\circ$) on the sky. Gaensler et al.\ \cite{ghs+05} re-analysed
archival LMC data taken with the ATCA, and extracted polarisation
and RMs for 292 background sources. The results, shown in
Figure~\ref{fig_lmc}, show that RMs are generally positive on the
eastern half of the galaxy, and negative on the western half.
Analysed in more detail, these RMs reveal a sinusoidal pattern as
a function of azimuth, implying a coherent, spiral, pattern in the
LMC's magnetic field, with a strength of about 1~$\mu$G \cite{ghs+05}.

The presence of this relatively strong, ordered, field is somewhat
surprising in the LMC. Standard turbulent dynamo theory requires
5--10~Gyr to amplify a weak primordial seed field to microgauss
levels, but the repeated tidal interactions between the LMC, Milky
Way and Small Magellanic Cloud should disrupt any field that might be
slowly built up through this process. The coherent field revealed
in Figure~\ref{fig_lmc} must have been amplified and organised rapidly,
in only a few hundred million years. One possibility is a cosmic
ray dynamo (e.g., \cite{hkol04}), which should thrive in the vigorous
starburst environment supplied by the LMC.


\section{Magnetism with the Square Kilometre Array}
\label{sec_ska}

\subsection{The Rotation Measure Grid}

\begin{figure}
\centerline{\psfig{file=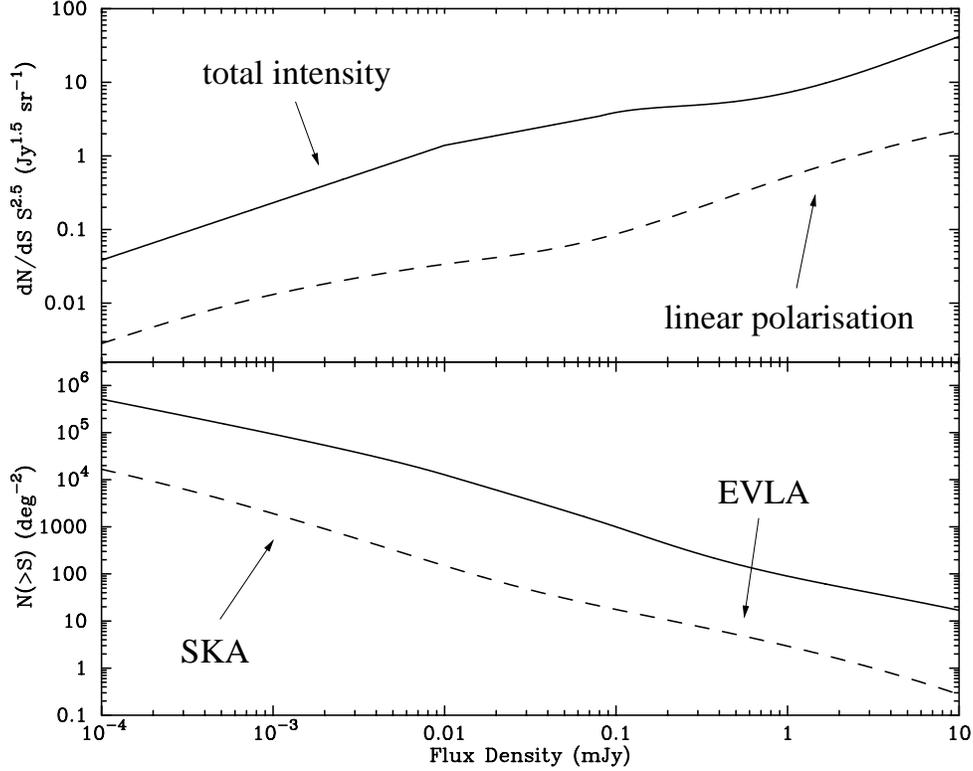,width=\textwidth}}
\caption{The predicted flux distribution of extragalactic radio
sources in both total intensity (solid lines) and in linear
polarisation (dashed lines), adapted from \cite{bg04}.  The upper
panel shows the differential source count distribution, while the
lower panel shows the integral distribution.  Approximate flux
limits for wide-field surveys with the EVLA and with the SKA are indicated.}
\label{fig_grid}
\end{figure}

\begin{figure}
\centerline{\psfig{file=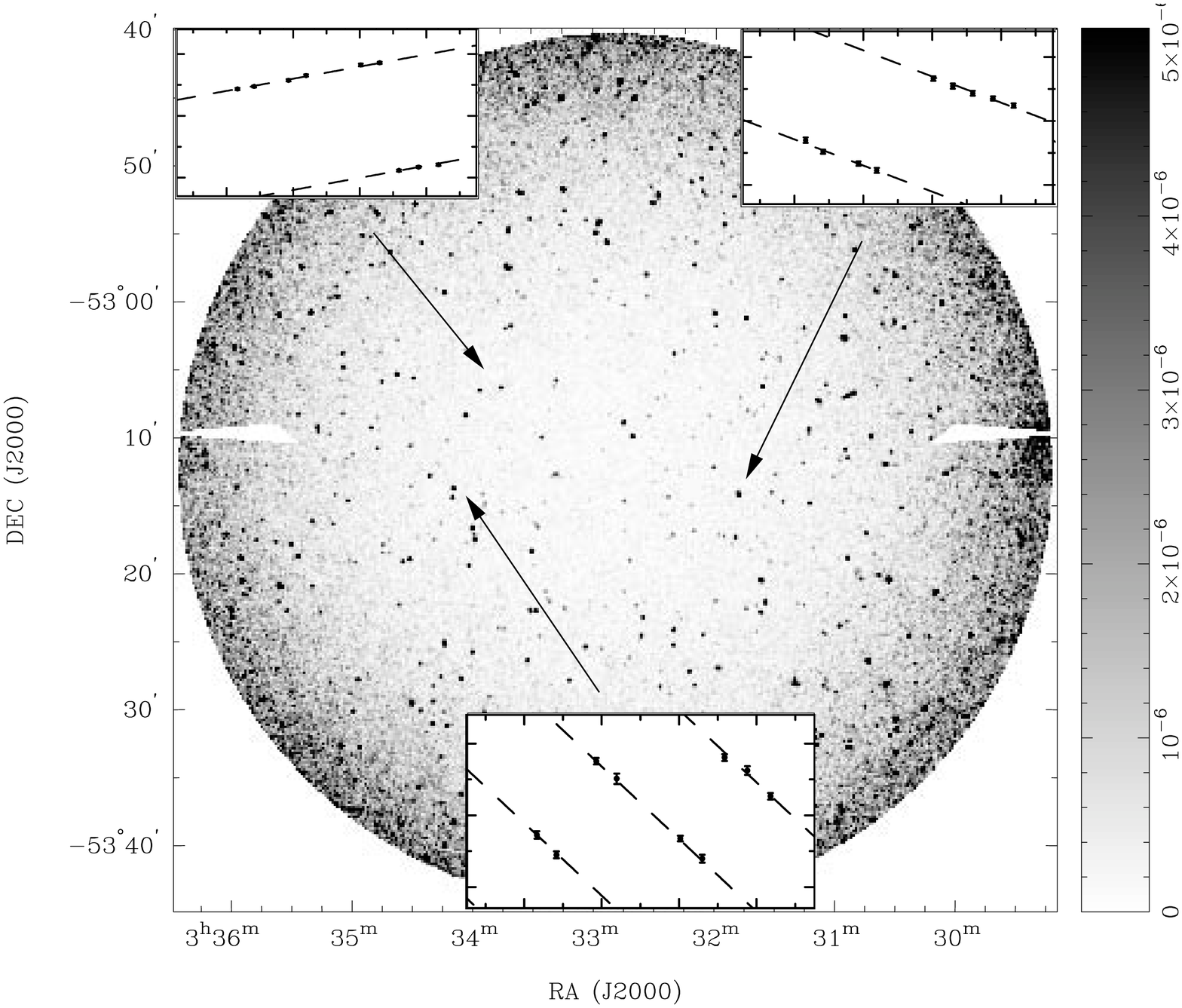,width=\textwidth}}
\caption{A simulation of a one-hour SKA observation in linear
polarisation, for a field of view of 1~deg$^2$. The angular resolution
is a few arcsec, and the gray scale is in units of $\mu$Jy. The
image was created by squaring the intensity values of an NVSS field
to generate a Ricean noise distribution, and then adjusting the
flux and spatial scales to simulate the SKA source density predicted
by Fig.~\protect\ref{fig_grid}.  The insets show some hypothesised
distributions of position angle vs.\ $\lambda^2$ for three sources
in the field, from which RMs can then be calculated.}
\label{fig_sim}
\end{figure}

The results presented in \S\ref{sec_mw} \& \S\ref{sec_lmc} can be
greatly expanded upon with a larger sample of RMs (see also Kronberg,
these proceedings). With the SKA, we envisage an all-sky ``rotation
measure grid'' \cite{bg04,gbf04}, which would be derived from a
1.4~GHz full-Stokes continuum survey. For an SKA field of view of
5~deg$^2$, six months of observing would result in an RMS sensitivity
of $\approx$$0.1$~$\mu$Jy~beam$^{-1}$. To estimate the yield of the
resulting RM grid, one needs to consider ``$\log N - \log P$'',
i.e., the differential source counts in linear polarisation, analogous
the usual $\log N - \log S$ function in total intensity \cite{bg04,tsg+07}.
This results in a distribution like that shown in Figure~\ref{fig_grid}.
While there are uncertainties in extrapolating to the low flux
levels expected for the SKA (see discussion by \cite{tsg+07}), we
can roughly predict that the RM grid should yield about $\sim10^8$
sightlines, with a typical spacing between measurements of $\sim1'$.
A simulation of the polarised sky as might be seen with the SKA
is shown in Figure~\ref{fig_sim}.
Such a data-base will provide a fantastic probe of all manner of
extended foreground sources, either individually (like the case of
the LMC) or as a statistical ensemble (see \cite{sab+08} and Arshakian
et al., these proceedings).

\subsection{The Magnetic Universe}

One of the main applications of the SKA's RM grid will be to study
the growth of galactic-scale magnetic fields as a function of cosmic
time.  The expectation is that the sightlines to distant extragalactic
sources should generally intersect one or more foreground galaxies,
as is seen in the Ly-$\alpha$ and Mg\,{\sc ii} absorption lines in
the optical spectra of quasars. Such intervenors should generate
an RM signature in the background source, and this signature should
potentially evolve with redshift. In particular, Equation~(\ref{eqn_rm2}),
if rewritten to take into account cosmological effects, contains a
$(1+z)^{-2}$ dilution term because of redshift of the emitted
radiation, but in some models can also contain co-moving terms with
dependencies $n_e \propto (1+z)^3$ and $B_\parallel \propto (1+z)^2$ \cite{wid02}.
The overall RM may then potentially evolve as rapidly as RM $\propto
(1+z)^3$, in which case we expect that filaments
and absorbers should begin to show an increasingly large RM at
higher $z$. If we can obtain a large sample of both RMs (from the
SKA) and accompanying redshifts (from the next generation of
photometric and spectroscopic optical surveys), we can apply a
variety of statistical tests  to the distribution of RM vs.\ $z$
to map the magnetic evolution of the Universe to $z \sim 3$
\cite{kol98,bbo99,kbm+08}.

\subsection{Magnetic Fields at $z > 5$}

There is already good evidence that microgauss-strength magnetic
fields exist out to redshifts $z \sim 1-2$ \cite{kbm+08,kpz92}.
If we can extend such data out to $z \ga 5$, we can potentially
obtain strong constraints on how large-scale magnetic fields
were created and then amplified. Such measurements
can be made by obtaining RMs for polarised sources at very high redshifts.

Indeed, radio emission has already been detected from two classes
of sources at $z > 6$: gamma-ray burst afterglows \cite{fck+06} and
quasars \cite{mbhw06}.  We currently lack the sensitivity to detect
linear polarisation and RMs from these objects, but such measurements
should be possible with the SKA. Furthermore, since the cosmic
microwave background is linearly polarised, deep observations at
the upper end of the SKA frequency range may be able to measure RMs
against it \cite{kl96b,shm04}.  Such an experiment, while challenging,
would probe the integrated Faraday rotation over almost all of the
Universe's history. In considering such measurements, it is important
to note that an RM measurement to a high-$z$ source does not provide
any direct constraints on high-$z$ magnetic fields on its own, since
the observed RM will also contain contributions from low-$z$
components of the sightline, and from the Milky Way foreground.
Once a high-$z$ RM data point has been obtained, deep radio and
optical observations of that field can yield a large number of  RMs
and redshifts for adjacent foreground objects (see Fig.~\ref{fig_sim}).
When the corresponding foreground contribution is then removed, the
high-$z$ magnetic field can be isolated studied.

\section{Conclusions}

Cosmic magnetism is a vigorous and rapidly developing field. What
makes this area particularly relevant for the SKA is that magnetic
fields at cosmological distances are uniquely probed at radio
wavelengths.  By studying the evolution of magnetic fields over the
Universe's history, we can simultaneously address a variety of major
topics in fundamental physics and astrophysics. This work also has
strong synergies with other astronomy and astroparticle experiments,
such as {\em Planck}, LSST, {\em JWST}, HESS and Auger \cite{ewvs06,apr07}.

In the coming years, a host of SKA pathfinders will begin to finally
reveal the depth and detail of the polarised sky \cite{rbb+06,jbb+07},
culminating in an exploration of the full Magnetic Universe with
the Square Kilometre Array.


\acknowledgments

I thank my various collaborators for their contributions to the
work reported here, in particular Jo-Anne Brown for providing the
material for several figures.  This work has been supported by the
Australian Research Council (grant FF0561298) and by the National
Science Foundation (grant AST-0307358).

\providecommand{\href}[2]{#2}\begingroup\raggedright\endgroup


\begin{thebibliography}{10}

\bibitem{zwe06}
E.~G. Zweibel, {\it Evolution of magnetic fields at high redshift},  {\em
  Astron. Nach.} {\bf 327} (2006) 505--509.

\bibitem{gr01}
D.~Grasso and H.~R. Rubinstein, {\it Magnetic fields in the early universe},
  {\em Phys. Rep.} {\bf 348} (2001) 163--266.

\bibitem{wid02}
L.~M. Widrow, {\it Origin of galactic and extragalactic magnetic fields},  {\em
  Reviews of Modern Physics} {\bf 74} (2002) 775--823.

\bibitem{mel80b}
D.~B. Melrose, {\em Plasma Astrophysics: Nonthermal Processes in Diffuse
  Magnetized Plasmas}.
\newblock Gordon \& Breach, New York, 1980.

\bibitem{gll05}
E.~M. de~Gouveia Dal~Pino, G.~Lugones and A.~Lazarian, {\em Magnetic Fields in
  the Universe}.
\newblock American Institute of Physics, Melville, New York, 2005.

\bibitem{zh97}
E.~G. {Zweibel} and C.~{Heiles}, {\it Magnetic fields in galaxies and beyond.},
   {\em Nature} {\bf 385} (1997) 131--136.

\bibitem{hei96}
C.~Heiles, {\it The local direction and curvature of the galactic magnetic
  field derived from starlight observations},  {\em ApJ} {\bf 462} (1996)
  316--325.

\bibitem{bh96}
R.~{Beck} and P.~{Hoernes}, {\it {Magnetic spiral arms in the galaxy NGC
  6946.}},  {\em Nature} {\bf 379} (1996) 47--49.

\bibitem{wkc+00}
D.~{Ward-Thompson}, J.~M. {Kirk}, R.~M. {Crutcher}, J.~S. {Greaves}, W.~S.
  {Holland} and P.~{Andr{\' e}}, {\it {First Observations of the Magnetic Field
  Geometry in Prestellar Cores}},  {\em ApJ} {\bf 537} (2000) L135--L138.

\bibitem{th86}
T.~H. Troland and C.~Heiles, {\it Interstellar magnetic field strengths and gas
  densities: {O}bservational and theoretical perspectives},  {\em ApJ} {\bf
  301} (1986) 339--345.

\bibitem{rs90}
M.~J. Reid and E.~M. Silverstein, {\it {OH} masers and the {G}alactic magnetic
  field},  {\em ApJ} {\bf 361} (1990) 483--486.

\bibitem{gmg98}
B.~M. Gaensler, R.~N. Manchester and A.~J. Green, {\it Radio continuum and {HI}
  observations of supernova remnant {G296.8--00.3}},  {\em MNRAS} {\bf 296}
  (1998) 813--823.

\bibitem{fer49}
E.~Fermi, {\it On the origin of the cosmic radiation},  {\em Phys. Rev.} {\bf
  75} (1949) 1169--1174.

\bibitem{dg49}
L.~Davis~Jr. and J.~L. Greenstein, {\it The polarization of starlight by
  interstellar dust particles in a galactic magnetic field},  {\em Phys. Rev.}
  {\bf 75} (1949) 1605.

\bibitem{man72}
R.~N. {Manchester}, {\it Pulsar rotation and dispersion measures and the
  galactic magnetic field},  {\em ApJ} {\bf 172} (1972) 43--52.

\bibitem{sk80}
M.~Simard-Normandin and P.~P. Kronberg, {\it Rotation measures and the
  {G}alactic magnetic field},  {\em ApJ} {\bf 242} (1980) 74--94.

\bibitem{tn80}
R.~C. Thomson and A.~H. Nelson, {\it The interpretation of pulsar rotation
  measures and the magnetic field of the {Galaxy}},  {\em MNRAS} {\bf 191}
  (1980) 863--870.

\bibitem{shu05}
A.~Shukurov, {\it Mesoscale magnetic structures in spiral galaxies},  in {\em
  Cosmic Magnetic Fields} (R.~Wielebinski and R.~Beck, eds.), (Springer, Berlin),
  pp.~113--135, 2005.

\bibitem{wck+04}
J.~M. {Weisberg}, J.~M. {Cordes}, B.~{Kuan}, K.~E. {Devine}, J.~T. {Green} and
  D.~C. {Backer}, {\it Arecibo 430 {MHz} pulsar polarimetry: {Faraday} rotation
  measures and morphological classifications},  {\em ApJS} {\bf 150} (2004)
  317--341.

\bibitem{hml+06}
J.~L. Han, R.~N. Manchester, A.~G. Lyne, G.~J. Qiao and W.~van Straten, {\it
  Pulsar rotation measures and the large-scale structure of {Galactic} magnetic
  field},  {\em ApJ} {\bf 642} (2006) 868--881.

\bibitem{val05}
J.~P. {Vall{\'e}e}, {\it Pulsar-based {Galactic} magnetic map: {A} large-scale
  clockwise magnetic field with an anticlockwise annulus},  {\em ApJ} {\bf 619}
  (2005) 297--305.

\bibitem{val02}
J.~P. Vall\'ee, {\it Metastudy of the spiral structure of our home {Galaxy}},
  {\em ApJ} {\bf 566} (2002) 261--266.

\bibitem{the63}
M.~Th\'evenot, {\em {Relations de Divers Voyages Curieux}}.
\newblock J.~Langlois, Paris, 1663.

\bibitem{mwkj03}
D.~Mitra, R.~Wielebinski, M.~Kramer and A.~Jessner, {\it The effect of {HII}
  regions on rotation measure of pulsars},  {\em A\&A} {\bf 398} (2003)
  993--1005.

\bibitem{cl02}
J.~M. {Cordes} and T.~J.~W. {Lazio}, {\it {NE2001. I. A New Model for the
  Galactic Distribution of Free Electrons and its Fluctuations}}, preprint
  (\href{http://arxiv.org/abs/astro-ph/0207156}{arXiv:astro-ph/0207156}).

\bibitem{btj03}
J.~C. {Brown}, A.~R. {Taylor} and B.~J. {Jackel}, {\it {Rotation Measures of
  Compact Sources in the Canadian Galactic Plane Survey}},  {\em ApJS} {\bf
  145} (2003) 213--223.

\bibitem{bhg+07}
J.~C. Brown, M.~Haverkorn, B.~M. Gaensler, A.~R. Taylor, N.~S. Bizunok, N.~M.
  McClure-Griffiths, J.~M. Dickey and A.~J. Green, {\it Rotation measures of
  extragalactic sources behind the southern {Galactic} plane: {New} insights
  into the large-scale magnetic field of the inner {Milky Way}},  {\em ApJ}
  {\bf 663} (2007) 258--266.

\bibitem{bt01}
J.~C. {Brown} and A.~R. {Taylor}, {\it {The Structure of the Magnetic Field in
  the Outer Galaxy from Rotation Measure Observations through the Disk}},  {\em
  ApJ} {\bf 563} (Dec., 2001) L31--L34.

\bibitem{ghs+05}
B.~M. Gaensler, M.~Haverkorn, L.~Staveley-Smith, J.~M. Dickey, N.~M.
  McClure-Griffiths, J.~R. Dickel and M.~Wolleben, {\it The magnetic field of
  the {Large Magellanic Cloud} revealed through {Faraday} rotation},  {\em
  Science} {\bf 307} (2005) 1610--1612.

\bibitem{hkol04}
M.~{Hanasz}, G.~{Kowal}, K.~{Otmianowska-Mazur} and H.~{Lesch}, {\it
  {Amplification of galactic magnetic fields by the cosmic-ray-driven dynamo}},
   {\em ApJ} {\bf 605} (2004) L33--L36.

\bibitem{bg04}
R.~Beck and B.~M. Gaensler, {\it Observations of magnetic fields in the {Milky
  Way} and in nearby galaxies with a {Square Kilometre Array}},  {\em New
  Astron. Rev.} {\bf 48} (2004) 1289--1304.

\bibitem{gbf04}
B.~M. Gaensler, R.~Beck and L.~Feretti, {\it The origin and evolution of cosmic
  magnetism},  {\em New Astron. Rev.} {\bf 48} (2004) 1003--1012.

\bibitem{tsg+07}
A.~R. Taylor, J.~M. Stil, J.~K. Grant, T.~L. Landecker, R.~Kothes, R.~Reid,
  A.~D. Gray, D.~Scott, P.~G. Martin, A.~Boothroyd, G.~Joncas, F.~J. Lockman,
  J.~English, A.~Sajina and J.~R. Bond, {\it Radio polarimetry of the {ELAIS
  N1} field: {Polarized} compact sources},  {\em ApJ} {\bf 666} (2007)
  201--211.

\bibitem{sab+08}
R.~Stepanov, T.~G. Arshakian, R.~Beck, P.~Frick and M.~Krause, {\it Magnetic
  field structures of galaxies derived from analysis of {Faraday} rotation
  measures, and perspectives for the {SKA}},  {\em A\&A} (2008), in press
  (\href{http://arxiv.org/abs/0711.1267}{arXiv:0711.1267}).

\bibitem{kol98}
T.~Kolatt, {\it Determination of the primordial magnetic field power spectrum
  by faraday rotation correlations},  {\em ApJ} {\bf 495} (1998) 564--579.

\bibitem{bbo99}
P.~Blasi, S.~Burles and A.~V. Olinto, {\it Cosmological magnetic field limits
  in an inhomogeneous universe},  {\em ApJ} {\bf 514} (1999) L79--L82.

\bibitem{kbm+08}
P.~P. Kronberg, M.~L. Bernet, F.~Miniati, S.~J. Lilly, M.~B. Short and D.~M.
  Higdon, {\it A global probe of cosmic magnetic fields to high redshifts},
  {\em ApJ} (2008), in press (\href{http://arxiv.org/abs/0712.0435}{arXiv:0712.0435}).

\bibitem{kpz92}
P.~P. {Kronberg}, J.~J. {Perry} and E.~L.~H. {Zukowski}, {\it {Discovery of
  extended Faraday rotation compatible with spiral structure in an intervening
  galaxy at $z = 0.395$: New observations of PKS 1229--021}},  {\em ApJ} {\bf
  387} (1992) 528--535.

\bibitem{fck+06}
D.~A. Frail, P.~B. Cameron, M.~Kasliwal, E.~Nakar, P.~A. Price, E.~Berger,
  A.~Gal-Yam, S.~R. Kulkarni, D.~B. Fox, A.~M. Soderberg, B.~P. Schmidt,
  E.~Ofek and S.~B. Cenko, {\it An energetic afterglow from a distant stellar
  explosion},  {\em ApJ} {\bf 646} (2006) L99--L102.

\bibitem{mbhw06}
I.~D. McGreer, R.~H. Becker, D.~J. Helfand and R.~L. White, {\it Discovery of a
  $z = 6.1$ radio-loud quasar in the {NOAO Deep Wide Field Survey}},  {\em AJ}
  {\bf 652} (2006) 157--162.

\bibitem{kl96b}
A.~Kosowsky and A.~Loeb, {\it Faraday rotation of microwave background
  polarization by a primordial magnetic field},  {\em ApJ} {\bf 469} (1996)
  1--6.

\bibitem{shm04}
C.~Sc\'occola, D.~Harari and S.~Mollerach, {\it B polarization of the {CMB}
  from {Faraday rotation}},  {\em Phys. Rev. D} {\bf 70} (2004) 063003.

\bibitem{ewvs06}
T.~A. En{\ss}lin, A.~Waelkens, C.~Vogt and A.~A. Schekochihin, {\it Future
  magnetic field studies using the {Planck} surveyor experiment},  {\em Astron.
  Nach.} {\bf 327} (2006) 626--631.

\bibitem{apr07}
A.~De~Angelis, M.~Persic and M.~Roncadelli, {\it Constraints on large-scale
  magnetic fields from the {Auger} results}, preprint,
\href{http://arxiv.org/abs/0711.3346}{arXiv:0711.3346}.

\bibitem{rbb+06}
H.~J.~A. R\"ottgering, B.~R., P.~D. Barthel, M.~P. van Haarlem,
G.~K. Miley,
  G.~K. Morganti and I.~Snellen, {\it {LOFAR --- Opening} up a new
  window on the {Universe}},  in
  \href{http://www-astro.physics.ox.ac.uk/LOSKA/CONFERENCE/PROCEEDINGS}{\em
  Cosmology, Galaxy Formation and Astroparticle Physics on the
  Pathway to the SKA} (H.-R. Kl\"ockner, S.~Rawlings, M.~Jarvis and
  A.~Taylor, eds.), 2006

\bibitem{jbb+07}
S.~{Johnston}, et al., {\it Science with the {Australian Square
Kilometre Array Pathfinder}},  {\em PASA} {\bf 24} (2007) 174--188.

\end{thebibliography}

\end{document}